\documentclass[a4paper,rmp,twocolumn,showkeys,superscriptaddress]{revtex4}

\usepackage{psfrag}
\usepackage[latin1]{inputenc}
\usepackage{graphicx}

\begin{document}

%\title{Language Networks: their structure, function and evolution}

\providecommand{\ICREA}{ICREA-Complex Systems Lab, Universitat Pompeu Fabra (GRIB), 
Dr Aiguader 80, 08003 Barcelona, Spain}
\providecommand{\SFI}{Santa Fe Institute, 1399 Hyde Park Road, Santa Fe NM 87501, USA}
\author{Bernat Corominas Murtra}
\affiliation{\ICREA}
\author{Sergi Valverde}
\affiliation{\ICREA}
\author{Ricard V. Solé}
\affiliation{\ICREA}
\affiliation{\SFI}

\title{Emergence of scale-free syntax networks}

\begin{abstract}
 { The  evolution of human language allowed  the efficient propagation
  of nongenetic information, thus  creating a new form of evolutionary
  change.  Language development in  children offers the opportunity of
  exploring  the emergence  of such  complex communication  system and
  provides a window to understanding the transition from protolanguage
  to language.  Here we present the first analysis of the emergence of
  syntax in terms of complex networks.  A previously unreported, sharp
  transition  is  shown  to occur  around  two  years  of age  from  a
  (pre-syntactic)  tree-like structure  to a  scale-free,  small world
  syntax network.  The nature of such transition supports the presence
  of an innate component pervading  the emergence of full syntax. This
  observation is difficult  to interpret in terms of  any simple model
  of  network  growth, thus  suggesting  that  some internal,  perhaps
  innate component  was at work.  We  explore this problem  by using a
  minimal model  that is able  to capture several  statistical traits.
  Our  results  provide evidence  for  adaptive  traits,  but it  also
  indicates that some key features of syntax might actually correspond
  to non-adaptive phenomena.}
\end{abstract}


\maketitle

\section{Introduction}

Human language stands as one  of the greatest transitions in evolution
\citep{Maynard:1995} but  its exact origins remain a  source of debate
and   is  considered   one  of   the  hardest   problems   in  science
\citep{Christiansen:2003,  Szamado:2006}.   Since  language  does  not
leave fossils,  our windows to  its evolution are limited  and require
extrapolation   from  different   sources   of  indirect   information
\citep{Bickerton:1990}.  Among  the relevant questions  to be answered
is the  leading mechanism driving language emergence:  Is language the
result of natural selection?  The use of population models under noisy
environments  is   consistent  with  such   selection-driven  scenario
\citep{Hurford:1989,  Nowak:1999,  Komarova:2004}.  

Other  approaches  have  suggested  the  importance  of  communicative
constraints  canalizing  the   possible  paths  followed  by  language
emergence  \citep{Ferrer:2003}.  Supporting such  communication system
there has to  be a symbolic system which it has  been for some authors
the  core question \citep{Deacon:1997}.   Finally, a  rather different
approach focuses on the evolution  of the {\em machine} that generates
human language.   The most remarkable  trait of such {\em  machine} is
the     possibility      of     generating     infinite     structures
\citep{Humboldt:1999,   Chomsky:1957,  Hauser:2002}  in   a  recursive
fashion.  The  evolution of such  ability alone, beyond  its potential
functionality,  is considered  by  some authors  the  main problem  in
language evolution \citep{Hauser:2002}.
\begin{figure}
\begin{center}
  \includegraphics[width=8.5 cm]{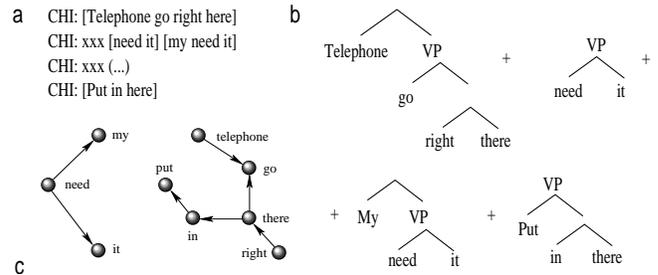}
\end{center}
\caption{Building  the networks  of Syntax  Acquisition. First  we
    identify   the  structures  in   child's  productions (a)  using  the
    lexico-thematic nature of early grammars \citep{Radford:1990}, see
    \citep{Bernat:2007}. Afterwards, a  basic constituency  analysis is
    performed (b) assuming  that the  semantically most relevant  item is
    the head of  the phrase and that the verb in  finite form (if any)
    is  the head  of the  sentence. Finally  (c) a  projection  of the
    constituent structure in a dependency graph is obtained.}
  \label{sample}
\end{figure}

An  alternative   approach  to   this  problem  considers   instead  a
non-adaptive view.   Roughly, language would be a  ``spandrel'' i.  e.
an  unselected  side-effect of  a  true adaptation  \citep{Gould:1979,
Gould:2002}.   The term  spandrel was  borrowed from  Architecture and
refers  to the  space between  two  arches or  between an  arch and  a
rectangular enclosure.  In the  context of evolution, a spandrel would
be a phenotypic characteristic that evolved as a side effect of a true
adaptation.   More precisely, the  features of  evolutionary spandrels
have  been summarized \citep{Sole:2006}  as follows  (a) they  are the
byproduct  (exaptation) of  building rules;  (b) they  have intrinsic,
well-defined, non-random features and (c) their structure reveals some
of the  underlying rules of system's  construction.  This non-adaptive
view   has   been   criticized   for   a  number   of   good   reasons
\citep{Dennet:1995}  but  remains as  an  important  component of  the
evolution  debate. Within the  context of  language evolution,  it has
been suggested that language would  have been a consequence of a large
brain,  with  neural  structures  formerly used  for  other  functions
\citep{Hauser:2002}.

Since there is no direct  trace of primitive communication systems, we
are forced to study this problem by means of indirect evidence, in the
hope that  ``no event happens in  the world without  leaving traces of
itself''\citep{Bickerton:1990}.   The remarkable  process  of language
acquisition  in children  is probably  the best  candidate for  such a
trace of  adaptation \citep{Bickerton:1990, Maynard:1995}.  Confronted
with the  surprising mastery of  complex grammar achieved  by children
over two years, some authors early concluded that an innate, hardwired
element   (a   language   acquisition   device)  must   be   at   work
\citep{Chomsky:1988, Pinker:1990, Pinker:1994}.   Children are able to
construct complex sentences  by properly using phonological, syntactic
and semantic rules  in spite that no one  teaches them.  Specifically,
they can  generate a virtually  infinite set of  grammatically correct
sentences in  spite that  they have been  exposed to a  rather limited
number  of input  examples.  Moreover,  although the  lexicon  shows a
monotonous growth as  new words are learned, the  pattern of change in
syntactic  organization  is strongly  nonlinear,  with a  well-defined
transitions from  babbling to a  fully, complex adult  grammar through
the  one  word and  two  words  stage  \citep{Radford:1990}.  

How  can children  acquire such  huge set  of rules?   Are  there some
specific, basic rules predefined as a part of the biological endowment
of  humans?   If  so,  some  mechanism of  language  acquisition  (the
universal  grammar) should  guide the  process.  In  this  way, models
assuming  a constrained  set of  accessible grammars  have  shown that
final  states (i.e., an  evolutionary stable  complex grammar)  can be
reached   under    a   limited   exposure   to    the   right   inputs
\citep{Komarova:2001,Niyogi:2006}.  However,  we cannot deny  the fact
that  important features of  the language  acquisition process  can be
obtained by  appealing only to general purpose  mechanisms of learning
\citep{Newport:1991, Elman:1993, MacWhinney:2005} or the importance of
pure   self-organization  in   the  structure   of  the   speech  code
\citep{Steels:1997,Oudeyer:2006}.   An integrated picture  should take
into account  the interaction of some predefined  grammar with general
purpose mechanisms of learning and code self-organization, structuring
human languages  as we know  today.  Under this view,  transition from
protogrammar to grammar would be  the result of an innovation of brain
organization rapidly predated for communication \citep{Hauser:2002}.

A quantitative  analysis of language  acquisition data is  a necessary
source of  validation of  different hypotheses about  language origins
and  organization. Indeed,  it is  well accepted  that  any reasonable
theory of language should be able  to explain how it is acquired. Here
we analyze  this problem  by using a  novel approximation  to language
acquisition based on a global,  network picture of syntax.  Instead of
following  the changes  associated  to lexicon  size  or counting  the
enumber of pairs  (or strings) of words, we rather  focus on how words
relate to each other and how  this defines a global graph of syntactic
links.  We focus our analysis in the presence of marked transitions in
the  global organization  of such  graphs.  As  shown below,  both the
tempo and mode of network  change seem consistent with the presence of
some predefined  hardware that is  triggered at some point  of child's
cognitive  development.  Furthermore,  we explore  this  conjecture by
means of an explicit model of  language network change that is able to
capture many (but not all)  features of syntax graphs.  The agreements
and  disagreements can  be interpreted  in terms  of  non-adaptive and
adaptive ingredients of language organization.

\section{Building syntactic Networks}
Language    acquisition    involves    several    well-known    stages
\citep{Radford:1990}. The first stage is the so-called {\em babbling},
where  only  single  phonemes   or  short  combinations  of  them  are
present. This stage  is followed by the {\em  Lexical spurt}, a sudden
lexical explosion where the child  begins to produce a large amount of
isolated words.  Such stage is  rapidly replaced by the {\em two words
stage},  where short  sentences of  two words  are produced.   In this
period,  we  do not  observe  the  presence  of functional  items  nor
inflectional morphology.   Later, close to  the two-years age,  we can
observe the {\em syntactic  spurt}, where more-than-two word sentences
are  produced.   The data set  studied  here  includes  a time  window
including  all the early,  key changes  in language  acquisition, from
non-grammatical to grammatical stages.
\begin{figure*}
\begin{center}
\includegraphics[width=14 cm]{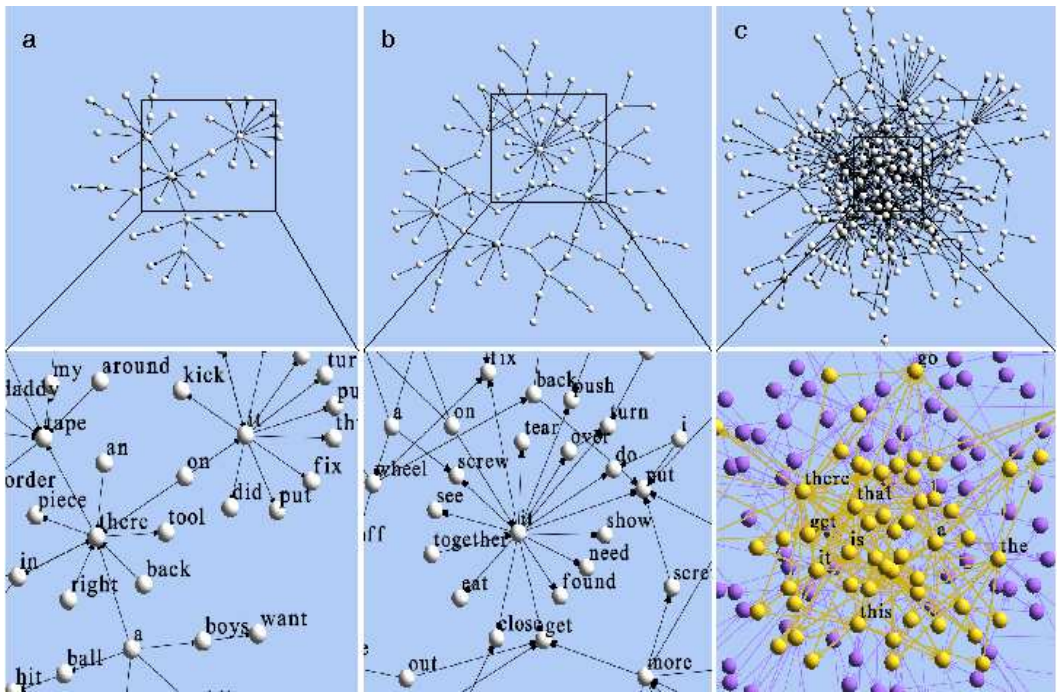}
\end{center}
\caption{Transitions from tree-like graphs to scale-free syntax graphs
through the  acquisition process. Here three snapshots  of the process
are shown, at (a) 25 months,  (b) 26 moths and (c) 28 months. Although
a   tree-like  structure   is  shown   to  be   present   through  the
pre-transition (a-b)  a scale-free,  much more connected  web suddenly
appears  afterward (c),  just two  months later.   The  lower pictures
indicate  how the hubs  are organized  and their  nature.  There  is a
critical change at the two-years age marked by a strong reorganization
of  the network.   Prior to  the transition,  semantically degenerated
elements (such as {\em it}) act as hubs.  Key words essential to adult
syntax are missing  in these early stages.  After  the transition, the
hubs change  from semantically degenerated to  functional items (i.e.,
{\em a} or  {\em the}).  In (f) we highlight the  core of this network
(the hubs and their links) using yellow nodes and edges.}
\label{3nets}
\end{figure*}

In this  paper we analyse  raw data obtained from  child's utterances,
from which we extract a global map of the pattern of the use syntactic
relations among words. In using this view, we look for the dynamics of
large-scale organization of the use of syntax. This can be achieved by
means  of complex  networks techniques,  by aggregating  all syntactic
relationships within a graph.  Recent studies have shown that networks
reveal   many   interesting    features   of   language   organization
\citep{Melcuck:1989,     Ferrer:2001,     Sigman:2002,    Ferrer:2004,
Hudson:2006,  Ke:2006} at different  levels.  These  studies uncovered
new  regularities in  language organization  but so  far none  of them
analyzed the  emergence of syntax through  language acquisition.  Here
we study in detail a  set of quantitative, experimental data involving
child utterances at different times of their development.

Formally, we define the  {\em syntax network} ${\cal G}={\cal G}({\cal
W},E)$ as  follows (see fig.\ref{sample}).   Using the lexicon  at any
given acquisition stage, we obtain  the collection of words $W_i (i=1,
..., N_w)$,  being every  word a node  $w_i \in  \cal G$.  There  is a
connection   between  two   given   words  provided   that  they   are
syntactically linked\footnote{Recall  that the  net is defined  as the
projection of  the constituency hierarchy.   Thus, the {\em  link} has
not     an     ontological     status     under    our     view     of
syntax\citep{Bernat:2007}}.  The  set of  links $E$ describes  all the
syntactic relationships  in the corpus.  For  every acquisition stage,
we  obtain a  syntactic  network  involving all  the  words and  their
syntactic  relationships.  The  structure of  syntax networks  will be
described  by means of  the {\em  adjacency matrix}  $A=[a_{ij}]$ with
$a_{ij}=1$  when there is  a link  between words  $w_i$ and  $w_j$ and
$a_{ij}=0$ otherwise.

Our corpora are extracted from a recorded session where a child speaks
with  adults spontaneously.   We  have collected  them  from the  {\em
CHILDES                                                       Database}
\citep{MacWhinney:2000}\footnote{http://talkbank.org}.   The  analysis
was   performed   using  the   {\em   Dependency  Grammar   Annotator}
\citep{Popescu:2003}.   Specifically, we choose  Peter's corpora  as a
particularly  representative and  complete  example \citep{Bloom:1974,
Bloom:1975}. Time intervals  are regular and the corpora  spans a time
window that  can be considered  large enough to  capture statistically
relevant properties.  Each corpus contains several conversations among
adult investigators  and the child.   However, the raw corpus  must be
parsed   in  order   to   construct  properly   defined  graphs.    In
\citep{Bernat:2007} we present a  detailed description of the criteria
and rules followed to pre-process  the raw data.  The main features of
the  parsing algorithm are  indicated in  fig.\ref{sample} and  can be
summarized as follows:

%%%%%%%%%%%%%%%%%%%%%%%%%%%%%%%%%%%%%%%%%%%%%%%%%%%%%%%%%%%%%%%%%%%%%%%%%
% Considerem que la construcció de la xarxa és al material suplementari %
%%%%%%%%%%%%%%%%%%%%%%%%%%%%%%%%%%%%%%%%%%%%%%%%%%%%%%%%%%%%%%%%%%%%%%%%%

\begin{enumerate}

\item
Select only  child's productions rejecting  imitations, onomatopoeia's
and undefined lexical items.

\item
Identify the {\em structures}, i.e., the minimal syntactic constructs.

\item
Among  the  selected  structures,  we  perform  a  basic  analysis  of
constituent structure, identifying the verb in finite form (if any) in
different phrases.

\item
Project  the constituent  structures into  lexical  dependencies. This
projection is close to  the one proposed by \citep{Hudson:2006} within
the  framework of the  network-based {\em  Word Grammar}\footnote{note
that the operation is reversible,  since can rebuild the tree from the
dependency relations}.

\item
Finally, we build  the graph by following the  dependency relations in
the  projection of  the syntactic  structures found  above. Dependency
relations allow us to construct  a syntax graph.
\end{enumerate}

With  this procedure, we  will obtain  a graph  for every  corpus. The
resulting graphs will be our object of study in the following section.

\section{Evolving syntax Networks}
\begin{figure}
\begin{center}
\includegraphics[width=8.5 cm]{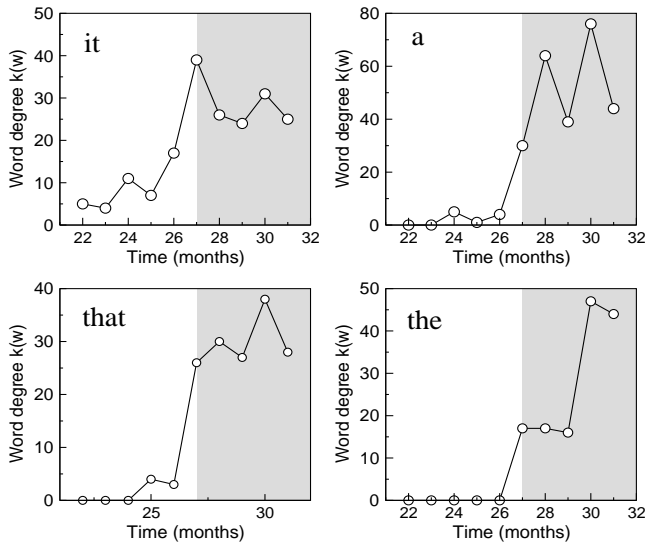}
\end{center}
\caption{Time evolution of  word degrees through language acquisition.
Here four  relevant words  have been chosen:  {\em it, a,  that, the}.
Their  degree  has  been  measured   in  each  corpus  and  display  a
well-defined  change  close  to  the  critical  age  of  $\approx  24$
months. Interestingly, {\em it} is  rapidly replaced by {\em a} as the
main hub as soon as purely functional words emerge. The gray are indicates 
the post-transition (syntactic) domain.}
\label{TimeWords}
\end{figure}
Here we analyze the  topological patterns displayed by syntax networks
at different stages of language acquisition. To our knowledge, this is
the first detailed analysis of  language network ontogeny so far.  The
resulting   sequence   exhibits    several   remarkable   traits.   In
fig. (\ref{3nets}) we show three examples of these networks.  At early
stages,  (fig.  \ref{3nets}a,b)  most  words are  isolated (not  shown
here) indicating a dominant  lack of word-word linkage. Isolated words
are not  shown in these  plots. For each  stage, we study  the largest
subset of connected  words or {\em giant component}  (GC).  The reason
for considering the largest connected component is that, from the very
beginning, the  GC is much  larger than any other  secondary connected
component  and  in  fact   the  system  shows  an  almost  all-or-none
separation between  isolated words and  those belonging to the  GC. In
other  words,  the  giant  component  captures  almost  all  word-word
relations.  By sampling  corpora at different times, we  obtain a time
series of connected networks ${\cal G}({\cal W}_T, E_T)$, where ${\cal
W}_T$ and $E_T$ are the set of words and links derived from the $T$-th
corpus, $T=1,...,11$.

\subsection{Global organization}

In  agreement  with  the  well-known presence  of  two  differentiated
regimes,  we  found  that  networks  before  the  two-year  transition
(fig.\ref{3nets}a-b) show a  tree-like organization, suddenly replaced
by  much larger, heterogeneous  networks (fig.\ref{3nets}c)  which are
very  similar to  adult syntactic  networks  \citep{Ferrer:2004}. This
abrupt change indicates  a global reorganization marked by  a shift in
grammar  structure.  This is  particularly obvious  in looking  to the
changes in the nature of  hubs before and after the transition. Highly
connected   words  in  the   pre-transition  stage   are  semantically
degenerated  lexical items, such  as {\em  it}. After  the transition,
hubs emerge as  functional items, such as {\em a}  or {\em the}. These
hubs were essentially nonexistent  in previous stages, as displayed in
fig.\ref{TimeWords}.

\subsection{Average degree}

A  first quantitative measure  is the  connectivity of  every element.
The number of links (or {\em degree} $k_i=k(w_i)$ of a given word $w_i
\in {\cal  W}$ gives  a measure of  the number of  different syntactic
relations in which such a word participates.  Figure (\ref{TimeWords})
shows the time series of $k$  for several relevant words.  All of them
display  a  sharp change  around  two-years  ($T=5$).   The gray  area
indicates  the presence of  syntactic organization  and words  such as
{\em a, the}  or {\em that} strongly increase  their presence and take
the  control   of  the  hub  structure  (compare   with  the  previous
figure). The advantage  of using degree as a  measure of the relevance
of a given word is  that this topological trait is largely independent
on its frequency of appearance.
\begin{figure}
\begin{center}
\includegraphics[width=8.5 cm]{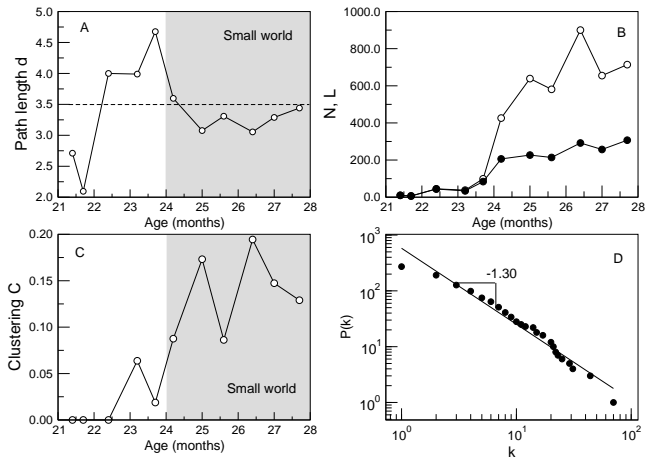}
\end{center}
\caption{Changes in  the structure of syntax networks  in children are
obtained by means of  several quantitative measures associated to the 
presence of small world and scale-free behavior. Here we display:
(a) the average path length $D_T$, (b) The number of words ($N_w$) and
links $L$ (c)  the clustering coefficient. As shown in  (a) and (c), a
small  world pattern  suddenly emerges  after an  age of  $\approx 24$
months. A  rapid transition from a  large $L$ and low  $C$ takes place
towards a small world network (with  low $D$ and high $C$).  After the
transition,   well-defined   scale-free   graphs,  with   $P(k)\propto
k^{-2.30}$, are observed (d).}
\label{Stats}
\end{figure}

\subsection{Small world development}

Two important measures allow  us to characterize the overall structure
of  these  graphs.   These  are  the average  path  length  $L_T$  and
clustering coefficient $C_T$ \citep{Watts:1998}.  The first measure is
defined  as  the average  $D_T=\left  <D_{min}(i,j)  \right >$,  where
$D_{min}(i,j)$ indicates  the length  of the shortest  path connecting
nodes $w_i$  and $w_j$.   The average is  performed over all  pairs of
words. Roughly speaking,  short path lengths means that  it is easy to
reach any given word $w_i$ starting from another arbitrary word $w_j$.
Small  path lengths  in sparse  networks  are often  an indication  of
efficient information  exchange.  The clustering  coefficient $C_T$ is
defined  as the probability  that two  words that  are neighbors  of a
given word are  also neighbors of each other (i.   e.  that a triangle
is formed). In order to estimate  $C_T$, we define for each word $w_i$
a  neighborhood   $\Gamma_i$.   Each   word  $w_j  \in   \Gamma_i$  is
syntactically related (at least once) with $w_i$ in a production.  The
words  in  $\Gamma_i$  can also  be  linked  to  each other,  and  the
clustering $C(\Gamma_i)$ is defined as
\begin{equation}
 C(\Gamma_i) = {1 \over k_i(k_i-1)}\sum_j \sum_{k\in\Gamma_i} a_{jk} 
\end{equation} 
\noindent
The  average clustering of  the $G_T$  network is  simply $C_T=\langle
C(\Gamma_i)\rangle$  i.e, the  average  over all  $w_i  \in W$.   Most
complex networks in  nature and technology are known  to be {\em small
words}, meaning that they have  short path lengths and high clustering
\citep{Watts:1998} Although language networks  have been shown to have
small world  structure \citep{Ferrer:2001, Steyvers:2005, Ferrer:2004,
Sigman:2002}  little  is known  about  how  it  emerges in  developing
systems.

Two  regimes in  language  acquisition  can be  also  observed in  the
evolution of the average path length fig.(4a).  It grows until reaches
a peak at the transition (where  the small word domain is indicated by
means of the grey area).  Interestingly, at $T=5$ the network displays
the highest number of words  for the pre-transition stage.  For $T>5$,
the   average  path   length   stabilizes  $D_T   \approx  3.5$   (see
fig. (\ref{Stats} b)).  The increasing  trend of $D_T$ in $T<5$ may be
an  indication that  combinatorial rules  are not  able to  manage the
increasing  complexity  of  the  lexicon.   In fig.(4b)  we  plot  the
corresponding  number of  words $N_T$  and links  $L_T$ of  the  GC as
filled and open circles, respectively.   We can see that the number of
connected  words that  belong  to  the GC  increases  in a  monotonous
fashion,  displaying a  weak jump  at the  age of  two.   However, the
number  of  links  (and  thus  the richness  of  syntactic  relations)
experiences a sharp change.

The  rapid increase  in the  number of  links indicates  a qualitative
change in  network properties  strongly tied to  the reduction  of the
average path length.  A similar  abrupt transition is observed for the
clustering  coefficient: In  the pre-transition  stage $C_T$  is small
(zero for  $T=1,2,3$).  After the transition, it  experiences a sudden
jump.  Both  $D_T$ and $C_T$ are  very similar to  the measured values
obtained    from     syntactic    graphs    from     written    corpus
\citep{Ferrer:2004}.

\subsection{Scale-free topology}

The small world behavior observed at the second phase is a consequence
of  the  heterogeneous distribution  of  links  in  the syntax  graph.
Specifically, we  measure the  degree distribution $P(k)$,  defined as
the  probability that  a node  has $k$  links. Our  syntactic networks
display  scale-free  degree  distributions $P(k)\propto  k^{-\gamma}$,
with  $\gamma\approx 2.3-2.5$.  Scale-free  webs are  characterized by
the presence of  a few elements (the hubs) having  a very large number
of  connections.    Such  heterogeneity   is  often  the   outcome  of
multiplicative  processes favouring  already  degree-rich elements  to
gain    further    links    \citep{Barabasi:1999,    Dorogovtsev:2001,
Dorogovtsev:2003}.

An example is shown  in fig.(\ref{Stats}d) where the cumulative degree
distribution, i.e:
\begin{equation}
P_{>}(k)=\int_k^{\infty}P(k)dk\sim k^{-\gamma+1}
\end{equation}
\noindent
is shown. The  fitting gives a scaling exponent  $\gamma \approx 2.3$,
also in agreement with adult studied corpora. They are responsible for
the very  short path  lengths and thus  for the  efficient information
transfer in complex networks. Moreover, relationships between hubs are
also   interesting:  the   syntax  graph   is   {\em  dissassortative}
\citep{Newman:2002}, meaning  that hubs tend to avoid  to be connected
among them  \citep{Ferrer:2004}. In  our networks, this  tendency also
experiences a sharp change close  to the transition domain (not shown)
thus indicating  that strong constraints emerge  strongly limiting the
syntactic linking between functional words.

\section{Modeling language acquisition}

\begin{figure}
\begin{center}
  \includegraphics[width=8.5 cm]{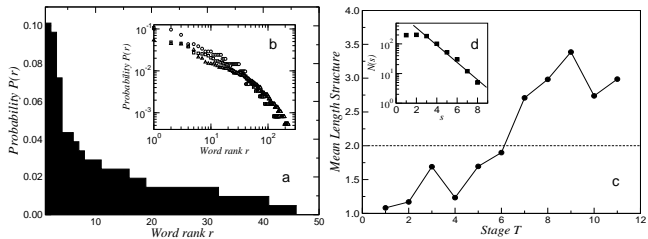}
\end{center}
\caption{Statistical  patterns  in  language  acquisition. In  (a)  an
example of  the rank-frequency distribution of lexical  items is shown
(here for  Peter's corpus  (see text)  at stage $T=2$  (1 year  and 10
months)).   The  inset (b)  displays  three  examples  of such  skewed
distributions in  log-log scale  for $T=2$ (circles),  $T=5$ (squares)
and  $T=8$  (triangles).  In  (c)  the  evolution  of mean  length  of
structure ($L$)  is displayed.  It  gives an estimate of  the (linear)
complexity  of the  productions  generated at  different stages.   The
dashed line indicates the two word production size. After stage $T=5$,
the $MSL$ ($\langle s \rangle$, in  the text) comes close to two and a
sharp change occurs.  In (d) we  also show an example of the frequency
distribution  $N(L)$  for these  productions  in  linear-log form  for
$T=5$.}
\label{SIMULNETS}
\end{figure}

We have described a pattern of change in syntax networks. The patterns
are nontrivial and quantitative. What  is their origin? Can we explain
them in terms of some  class of self-organization (SO) model? Are they
instead  associated to  some internal,  hardwired component?   Here we
present a  new model  of network evolution  that tries to  capture the
observed  changes  and  provides  tentative answers  to  the  previous
questions.

\subsection{Simple SO graph growth models}

We explored  several types of SO models  without success.  Appropriate
models  should be  able  to  generate: (a)  sharp  changes in  network
connectivity and  (b) scale-free  graphs as the  final outcome  of the
process.  In  relation to the  sudden shift, it  is well known  that a
sharp change in graph connectivity  occurs when we add links at random
between pairs of  nodes until a critical ratio  of links against nodes
is reached \citep{Erdos:1959, Bollobas:2001}.   Starting from a set of
$N$  isolated elements,  once the  number of  links $L$  is  such that
$p\equiv  L/N \approx  1$, we  observe a  qualitative change  in graph
structure, from  a set of small,  separated graphs ($p<1$)  to a graph
structure displaying  a giant  component ($p>1$) with  a comparatively
small  number of isolated  subgraphs. This  type of  {\em percolation}
model   has   been   widely    used   within   the   context   of   SO
\citep{Kauffman:1993, Sole:2001}.  Unfortunately, such a transition is
not satisfactory to explain our data,  since (a) it gives graph with a
Poissonian degree distribution  \citep{Bollobas:2001}, i.e.
\begin{equation}
P(k)\approx \frac{\langle k \rangle^ke^{-k}}{k!}
\end{equation}
\noindent 
and  (b) there is  no sharp  separation between  isolated nodes  and a
single connected graph, but  instead many subgraphs of different sizes
are observed.

Other  models  instead  consider  growing  graphs  using  preferential
attachment         rules        \citep{Barabasi:1999,Dorogovtsev:2001,
Dorogovtsev:2003}. In these models the number of nodes grows by adding
new  ones  which   tends  to  link  with  those   having  the  largest
connectivity (a  rich-gets-richer mechanism).  Under a broad  range of
conditions these amplification  mechanisms generate scale-free graphs.
However, the  multiplicative process does  not lead to  any particular
type of  transition phenomenon.  The status of  hubs remains  the same
(they just  win additional links).  Actually, well-defined predictions
can be made,  indicating that the degree of the  hubs scales with time
in a power-law form \citep{Barabasi:1999,Dorogovtsev:2001}.

Although many  possible combinations of the  previous model approaches
can be  considered, we  have found that  the simultaneous  presence of
both  scale-free structure  emerging  on top  of  a tree  and a  phase
transition  between  both  is  not  possible.  In  order  to  properly
represent the  dynamics of our  network, a data-driven  approach seems
necessary.

\begin{figure*}
  \includegraphics[width=13 cm]{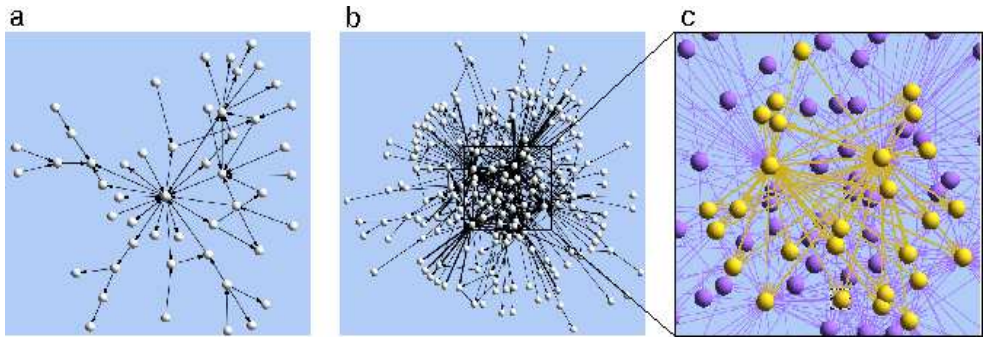}
\caption{Sudden  changes  in network  organization  from the  language
acquisition model (see  text).  In (a) and (b)  we display the largest
subgraph before  (c) and right  after (b) the transition.   The graphs
share  the  basic  change  from  tree-like  to  scale-free  structure,
although exhibit higher clustering  coefficients.  In (c) a blow-up of
(b) is shown, indicating the presence of a few hubs that are connected
among them both directly and through secondary connectors.}
\label{SIMULNETS2}
\end{figure*}

\subsection{Network growth model and analysis}

In order  to reproduce  the observed trends,  we have developed  a new
model of  network evolution.  The idea  is to describe  the process of
network  growth  without  predefined  syntactic rules.   We  make  the
simplistic  assumption  that word  interaction  only  depends on  word
frequency following Zipf's law. In this context, it has been suggested
that  Zipf's law  might be  the optimal  distribution  compatible with
efficient     communication     \citep{Harremoes:2001,    Ferrer:2003,
Ferrer:2005, Sole:2005}.  If no  internal mechanisms are at work, then
our model  should be able to  capture most traits of  the evolution of
syntax.

In  order to  develop the  model, a  new measure,  close to  the usual
$MLU$\footnote{The  {\em MLU} is  the {\em  Mean Length  of Utterance}
i.e.  the average  length of a child's utterances,  measured in either
words or morphemes.  }used in  linguistics, must be defined.  The {\em
structure  length}  of the  $i$-th  structured  production ($s_i$)  is
measured  by counting  the number  of  words that  participate in  the
$i$-th syntactic structure.  In our previous example (see figure 1) we
had  $4$   structures,  of  sizes  $|s_1|=4,   |s_2|=2,  |s_3|=2$  and
$|s_4|=3$.  Its  average, the {\em  Mean Structure Length,  $\langle s
\rangle$} is $\langle s \rangle = 2.75$.  In fig.  (\ref{SIMULNETS}-c)
we can  see how the  $MSL$ evolves over  time.  The frequency  of $s$,
$p(s)$ was  also measured and  was found to decay  exponentially, with
$p(s)\propto e^{-|s|/\gamma}$, with $\gamma=1.40$ in this specific set
of data (fig.  (\ref{SIMULNETS}-d)).   We can connect the two previous
through
\begin{equation}
\langle  s   \rangle=\frac{1}{Q}\sum_s  s  e^{-|s|/\gamma}
\end{equation}
where $Q$ is defined as the normalization constant:
\begin{equation}
Q=\sum_s e^{-|s|/\gamma}
\end{equation}
In the five first corpora, $\langle s \rangle<2$.
Beyond this stage,  it rapidly grows with $\langle  s \rangle>2$, (see
fig. (\ref{SIMULNETS}-b)).

We  incorpore to  the  data-driven model  our  knowledge on  structure
lengths.   We first  construct, for  each corpus,  a  random syntactic
network that  shares the statistics of word  frequencies and structure
lengths  of  the  corresponding  data  set.  Such  a  measure  can  be
interpreted, in  cognitive terms, as  some kind of working  memory and
might   be   the    footprint   of   some   maturational   constraints
\citep{Newport:1991, Elman:1993}.
\begin{figure*}
\begin{center}
\includegraphics[width=12.5 cm]{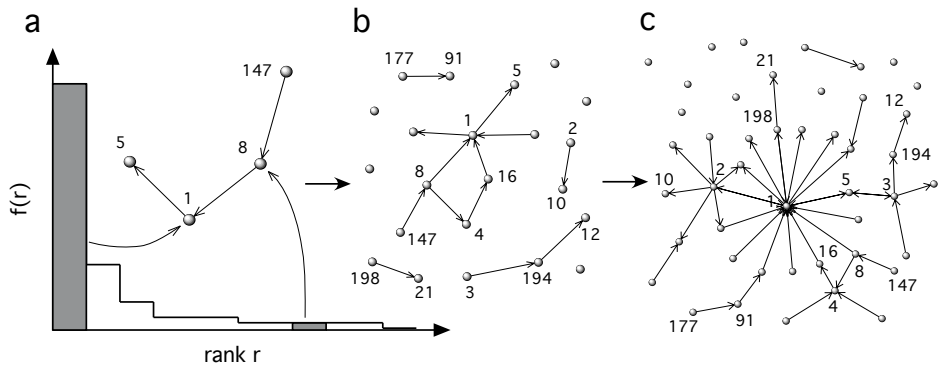}
\end{center}
\caption{Algorithm  for  network  growth.   The model  uses  as  input
information a Zipf's distribution  of ``words'' and the probability to
find a structure  of size $s$ in a given  corpus, $p_T(s)$.  Each step
we  choose $s$  words  from the  list,  each word  with a  probability
proportional to  their frequency.  A link is  then established between
two successive  words generating an unstructured string  of $s$ nodes.
We repeat the  process a number of times and we  aggregate in a global
graph all  the obtained strings.   $p_T(s)$ can be interpreted  as the
footprint  of a  kind of  working memory,  and follows  an exponential
distribution (As shown in fig. (\ref{SIMULNETS})) }
\label{Algorithm}
\end{figure*}
For  simplicity, we  assume  that  the probability  of  the i-th  most
frequent word is a scaling law:
\begin{equation}
  p_w(i) = {1 \over Z} i^{-\beta}
\end{equation}
\noindent
with  $1  \le  i  \le  N_w(T)$,  $\beta \approx  1$  and  $Z$  is  the
normalization constant:
\begin{equation}
  Z = \sum_{i=1}^{N_w(T)}\left(\frac{1}{i}\right)^{\beta}
\end{equation}
\noindent
(notice that $Z$ depends on lexicon size, $N_w(T)$, which grows slowly
at this  stage).  However,  the actual word  frequency is  affected by
other corpus features. In particular, our corpora are highly redundant
with many  duplicated structures but  we build our nets  ignoring such
redundancies, since  we are interested in the  topological patterns of
use.  For  every corpus  $T$  with  $N_s(T)$  distinct structures,  we
compute the distribution  of structure lengths $p_T(s)$, $1  \le T \le
11$.  From  $N_w(T)$, $p_w(i)$, $N_s(T)$  and $p_T(s)$, we  generate a
random  syntactic  network for  every  stage $1  \le  T  \le 11$  (see
fig.(\ref{Algorithm})).   Given  a  lexicon with  $N_w(T)$  different
items, labeled as $a_1...a_{N_w(T)}$ the model algorithm goes as 
follows:

\begin{enumerate}

\item
Generate a random positive integer $s$ with probability $p_T(s)$.

\item
Choose  $s$ different  ``words'' from  the  lexicon, $a_k^1,...,a_j^s$
each  word with  probability $p(a_i)\propto  i^{-\beta}$,  with $\beta
\approx 1$.

\item
Trace  an arc  between every  two successive  words thus  generating a
unstructured string of  $s$ nodes.

\item
Repeat (1), (2) and (3) until $N_s(T)$ structures are generated.

\item
Aggregate all the obtained strings in a single, global graph.

\end{enumerate}

\begin{figure}
\begin{center}
  \includegraphics[width=8.5 cm]{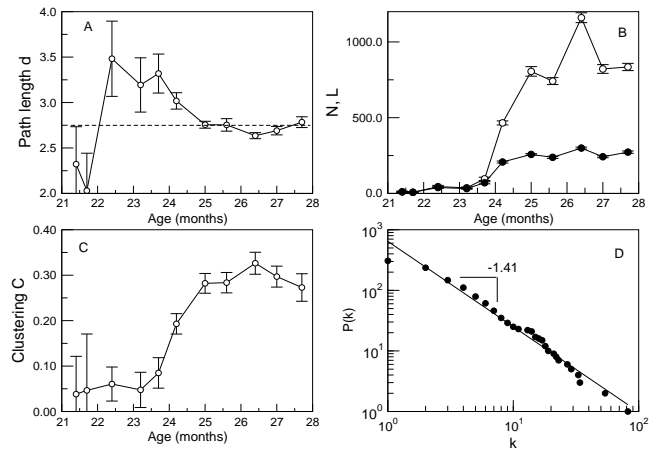}
\end{center}
\caption{Changes in  the structure  of syntax model  networks -compare
with fig.(\ref{Stats}). Here we show: (a) the average path length $L$,
(b) the  number of  links ($L$)  and lexical items  ($N$) and  (c) the
clustering  coefficient   $C$.   An   example  of  the   resulting  SF
distributions is also shown in (d).}
\label{StatModel}
\end{figure}

In  spite of the  small number  of assumptions  made, the  above model
reproduces many  of the topological traits observed  in real networks.
To begin with, we clearly observe the sudden transition from tree-like
networks   to    scale-free   networks   (see   fig.\ref{SIMULNETS2}).
Furthermore,  typical network properties,  such as  clustering, degree
distribution or path  lenghts seem to fit real  data successfully (see
fig.   (\ref{StatModel})).   The very  good  agreement between  global
patterns  of network  topology is  remarkable given  the lack  of true
syntax. It indicates that some essential properties of syntax networks
come  ``for free''.   In other  words, both  the small  world  and the
scale-free architecture of syntax  graphs would be spandrels: although
these type  of networks provide  important advantages (such  as highly
efficient and  robust network interactions) they would  be a byproduct
of  Zipf's law and  increased neural  complexity.  These  results thus
support the non-adaptive nature of language evolution.

However, particularly  beyond the  transition, a detailed  analysis is
able to find important  deviations between data and model predictions.
This  becomes  specially  clear  by  looking  at  small  subgraphs  of
connected words. Studying small size subgraphs allows to explore local
correlations among units. Such correlations are likely to be closer to
the underlying  rules of network construction, since  they are limited
specificaly to direct node-node relations and their frequency. We have
found  that the  subgraph census  reveals strong  deviations  from the
model due to the presence of grammatical constraints, i.e, non-trivial
rules to build the strings.

\begin{figure}
\begin{center}
\includegraphics[width=8.5 cm]{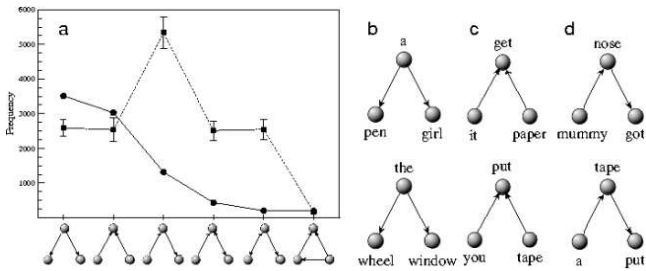}
\end{center}
\caption{Subgraph census  plot for  both real (circles)  and simulated
 (squares)  networks.   As   we  can  see  in  (a),   there  exist  an
 overabundance  of   the  first  two  subgraphs   due  to  grammatical
 restrictions on the  role of the syntactic head  (see text).  (b) and
 (c) are  an example  of the  kind of nodes  that participate  in such
 small subgraphs.  Beyond this two subgraphs, we find a sharp decay in
 its abundance against,  compared with the model.  This  is due to the
 fact that the  third studied motif (d) should be  abundant (as in the
 model).}
\label{Motifs}
\end{figure}

In figure (\ref{Motifs}) we display the so-called subgraph census plot
\citep{Holland:1970,  Wasserman:1994}  for  both  real  (circles)  and
simulated  (squares)  networks.   Here  the  frequencies  of  observed
subgraphs of size three are  shown ordered in decreasing order for the
real case. For  the simulated networks, we have  averaged the subgraph
frequencies  over  $50$  replicas.   Several obvious  differences  are
observed between both  censuses. The deviations are mainly  due to the
hierarchical relations that display a typical syntactic structure, and
to the fact that lexical items  tend to play the same specific role in
different structures (see fig.\ref{Motifs}b-d).  Specifically, we find
that the asymetries in syntactic relations induce the overabundance of
certain  subgraphs and  constrain the  presence of  others.  Specially
relevant is the  low value of third type  of subgraph, confronted with
the  model  prediction.  This  deviation   can  be  due  to  the  {\em
organizing}  role  of functional  words  (mainly  out-degree hubs)  in
grammar.  Indeed,  coherently with  this interpretation, we  find that
the  first type  of subgraph  (related with  out-degree hubs)  is more
abundant than the model prediction.

The second  interesting deviation is  given by the changing  status of
hubs. As  previously described, in  the prefunctional period  hubs are
semantically degenerated words, such as {\em that, it}, whereas beyond
the transition hubs are functional words. This observation seems to be
coherent  with a  recently  proposal to  understand  the emergence  of
functional  items in child  grammars. In  short, a  pure articulattory
strategy  introduces a  new sound  (mainly  the $a$)  that is  rapidly
predated   by  the  syntactic   system  when   it  is   mature  enough
\citep{Veneziano:2000}.   This would  imply  a reuse  of an  existing,
phonetical  element and  would explain  the astonishing  increasing of
appearance that they experience. If we follow the changes in number of
links displayed by the hubs  in the simulated system, no such exchange
is ever observed.  Instead,  their degree simply keeps growing through
the process (not shown).
\section{Discussion}  
Our study  reveals two clearly  differentiated behaviors in  the early
stages of language acquisition.   Rules governing both grammatical and
global behavior seem to be qualitatively and quantitatively different.
Could we explain the transition  in terms of self-organizing or purely
external-driven  mechanism?  Clearly not,  given the  special features
exhibited by our evolving webs,  not shared by {\em any} current model
of evolving networks \citep{Dorogovtsev:2001, Dorogovtsev:2003}.
Beyond the transition, some features diverge dramatically from the pre
transition graph,  particularly the changing  role of the  hubs.  Such
features  cannot   be  explained   from  external  factors   (such  as
communication constraints  among individuals). Instead,  it seems tied
to changes in the internal  machinery of grammar. The sharp transition
from small  tree-like graphs to  much larger scale-free nets,  and the
sudden  change  of  the nature  of  hubs  are  the footprints  of  the
emergence of  new, powerful rules of exploration  of the combinatorial
space,  i.e., the  emergence  of  full adult  syntax.   This seems  to
support    the    hypotheses    suggested    by    Hauser    et    al.
\citep{Hauser:2002}; see also \citep{Nowak:1999}.

Furthermore,  we   have  presented   a  novel  approach   to  language
acquisition  based  on a  simple,  data-driven  model. Previous  model
approaches  based on self-organization  cannot reproduce  the observed
patterns of  change displayed by syntax  graphs. Our main  goal was to
explore the potential roles of adaptive versus non-adaptive components
in shaping syntax networks as they  change in time.  The model is able
to reproduce some fundamental  traits.  Specifically we find that: (a)
the  global  architecture  of   syntactic  nets  obtained  during  the
acquisition process can be reproduced by using a combination of Zipf's
law and  assuming a growing  working memory and (b)  strong deviations
are observed when looking at the behavior of hubs and the distribution
of  subgraph  abundances.   Such  disagreements  cannot  be  fixed  by
additional rules. Instead, they  indicate the presence of some innate,
hard-wired  component  related with  the  combinatorial  power of  the
underlying grammatical  rules that is  triggered at some point  of the
child's cognitive  development. Our study  supports the view  that the
topological  organization  of  syntactic  networks is  a  spandrel,  a
byproduct  of communication  and neural  constraints.  But  the marked
differences found here cannot be  reduced to such scenario and need to
be  of   adaptive  nature.   Furthermore,  our   analysis  provides  a
quantitative argument to go forward beyond statistics in the search of
fundamental   rules   of  syntax,   as   it   was   early  argued   in
\citep{Miller:1963}.

A  further  line of  research  should  extend  the analysis  to  other
(typologically  different) languages  and  clarify the  nature of  the
innovation. Preliminary work  using three different european languages
supports our previous results (Corominas-Murtra et al {\em unpublished
work}).  Moreover, modeling  the transitions  from finite  grammars to
unbounded    ones   by    means   of    connectionist   approximations
\citep{Szathmary:2007} could shed  light on the neuronal prerequisites
canalizing the  acquisition process towards a  fully developed grammar
as described and measured by our network approach.

\section{Acknowledgements} 
The authors  thank Guy Montag  and the members  of the CSL  for useful
discussions.   We  also   acknowledge  Liliana  Tolchinsky  and  Joana
Rossell\'o  for  helpful comments  on  theory  of syntax  acquisition.
Finally,  we acknowledge Maria  Farriols i  Valldaura for  her support
during the whole  process of this work.  This  work has been supported
by  grants FIS2004-0542,  IST-FET  ECAGENTS, project  of the  European
Community  founded  under  EU   R\&D  contract  01194,  the  McDonnell
foundation (RVS) and by the Santa Fe Institute.

%\bibliography{lang}
%\bibliographystyle{natbib}

\end{document}